\documentclass[aps,pre,twocolumn,groupedaddress,showpacs]{revtex4}
\usepackage{graphicx}
\usepackage{amssymb}

\begin{document}

\title{Statistics of bubble rearrangement dynamics in a coarsening foam}

\author{A.S. Gittings$^1$ and D.J. Durian$^{1,2}$}
\affiliation{$^1$Department of Physics \& Astronomy, University of
California, Los Angeles, CA 90095, USA \\
$^2$Department of Physics \& Astronomy, University of
Pennsylvania, Philadelphia, PA 19104, USA}

\date{\today}

\begin{abstract}
We use speckle-visibility spectroscopy to measure the time-dependence of bubble rearrangement events that are driven by coarsening in an aqueous foam.  This technique gives the time-trace for the average scattering site speed within a prescribed volume of the sample.  Results are analyzed in terms of distributions of event times, event speeds, and event displacements.  The distribution of rest times between successive events is also measured;  comparison with diffusing-wave spectroscopy results shows that the spatial structure of a typical event consists of a core of only a few bubbles which undergo topology change plus a surrounding halo of bubbles which shift by an amount that decays to one wavelength at four to five bubbles away.  No correlations are found between the durations, speeds, and rest times between successive events.
\end{abstract}

\pacs{}

\maketitle

%=========================================================================================
\section{Introduction}

Aqueous foams consist of gas bubbles dispersed in a surfactant solution~\cite{WeaireBook99, KirkOthmerFoam04, StJalmesSM06}.  With proper formulation, the soap films between neighboring bubbles are completely stable against rupture.  Therefore, the bubbles may be considered as packing units of fixed but deformable volume.  If the foam is sufficiently dry, such that the bubbles occupy a volume fraction greater than random-close packing for spheres, about 0.64, then the foam will exhibit a solid-like character with elastic moduli set by the ratio of surface tension to bubble size~\cite{AndyARFM88, hohlerJPCM05}.  Since the bubbles are macroscopic, the corresponding energy scale of modulus times bubble volume is many orders of magnitude larger than $k_{\rm B}T$.  Microscopically, the bubbles are therefore jammed, i.e.\ locked into a mechanically stable packing configuration where neighbors cannot change due to thermal motion~\cite{LiuNagelBook}.  For bubbles to change neighbors, and for foams thereby to unjam and display a liquid-like character, requires an input of energy.  This may be supplied through an externally imposed shear deformation.  The energy for rearrangement may also be supplied through time evolution by coarsening, in which gas slowly diffuses from smaller to larger bubbles so that the total interfacial area decreases~\cite{Stavans93}.  For very dry foams, rearrangements may be considered in terms of topology changes ``of the first kind'', or T1 events, in the connected network of soap films~\cite{WeaireBook99}.

The nature of microscopic bubble rearrangements and their connection to macroscopic foam rheology have thus been topics of fundamental interest.  For bulk foams, Diffusing-Wave Spectroscopy (DWS) has emerged as an important tool~\cite{DJDsci91, DJDpra91, earnshawPRE94, AnthonyPRL95, hohlerPRL97, AnthonyJCIS99, hohlerPRL01}.  This is a dynamic light scattering technique for opaque samples~\cite{Weitz1993, Maret97}, whereby far-field intensity fluctuations at an area comparable to speckle size are measured and related to the absolute normalized electric field autocorrelation function, $g_1(\tau)$.  The relative motion of scattering sites broadens the power-spectrum of the light, given by the Fourier transform of $g_1(\tau)\exp(i\omega_\circ\tau)$ where $\omega_\circ$ is the incident frequency, and causes the speckle pattern to fluctuate.  Thus the functional form and decay rate of $g_1(\tau)$ vs $\tau$ give information on the nature and the rate of the scattering site motion, respectively.  For coarsening foams \cite{DJDsci91, DJDpra91}, and for slowly sheared foams where the bubbles lock into place between rearrangements, \cite{AnthonyJCIS99}, the contribution to $g_1(\tau)$ per scattering site is exponential in $\tau$ with a decay time scale $\tau_o$ set by the average time between successive rearrangement events at a site.  For rapidly sheared foams, where dissipative forces dominate surface tension forces so that the bubbles continuously flow and never lock into locally stable packing configurations, the contribution to $g_1(\tau)$ per scattering site is exponential in $\tau^2$ with a decay time scale set by the strain rate and how long it takes for adjacent scattering sites to convect apart by the wavelength of light \cite{AnthonyJCIS99}.

While DWS thus reveals microscopic information through the shape and decay rate of the field autocorrelation, it cannot capture several key features of the bubble rearrangement dynamics.  For example, it gives the average time between events, but it cannot give the distribution.  It cannot give the number of bubbles involved in an event.  It cannot probe the bubble-size changes during coarsening and the resulting readjustment of the packing.  Furthermore it cannot give the speed of bubble motion or the duration of the events.  The latter is particularly important because it sets the crossover strainrate below which the foam exhibits discrete rearrangements and a solid-like rheology and above which it exhibits continuous microscopic flow and a liquid-like rheology \cite{ParkPRL94, AnthonyJCIS99, AnthonyPRL03}.  These limitations of DWS arise because the signal is acquired by time-averaging the speckle fluctuation statistics and also because the scattering volume must be large compared to the event volume.

In this paper we apply a time-resolved dynamic light scattering technique called Speckle-Visibility Spectroscopy (SVS) \cite{Dixon2003, Bandyopadhyay2005} to a coarsening foam.  The advantages over DWS are that the scattering volume need not be large compared to the event volume, and that speckle fluctuations are probed instantaneously by a single exposure of a CCD camera.  From a time-series of such exposures, SVS allows the motion of individual rearrangement events to be examined directly.  After reviewing details of this method, we present results on the detailed statistics of bubble rearrangements.  This includes the full distribution of the time between successive events within the scattering volume, and by comparison with DWS results, an inference on average event size. It also includes the full distributions of event durations and peak bubble speeds.

%=========================================================================================
\section{Experimental methods}

\subsubsection{Foam}

The experimental foaming system is Gillette Foamy Regular shaving cream, as in many prior studies dating back to Refs.~\cite{DJDsci91, DJDpra91} and continuing as recently as Refs.~\cite{AnthonyPRL03,hohlerPRL04}.  The surfactants in aqueous solution are stearic acid and triethanolamine, plus other additives, giving a viscosity of $\mu=1.8$~cP and a surface tension of $\gamma=24$~erg/cm$^3$ \cite{AnthonyJCIS99}.  Though no doubt the exact formulation evolves according to market influences, we notice no difference with any precedents~\cite{Dates}.  Since this foam is produced by the aerosol method, the bubble sizes are smaller than possible with a mechanical mixing method.  The advantage is that, therefore, it is possible to let the system age {\it in-situ} and coarsen significantly prior to data collection and thus to achieve a reproducible history-independent distribution of bubble sizes.  Here we inject the foam into a rectangular glass cell with inner dimensions $0.7 \times 3 \times 10~{\rm cm}^3$, and seal.  Data are collected throughout $4.4-7.1$~hours after injection.  During this period there is no drainage or film rupture, the liquid fraction is constant at about 0.08, and the average bubble diameter grows from 100 to 120~microns (roughly linear in time, since the interval is small, even though the actual growth is a power-law with exponent near $1/2$) \cite{DJDpra91}.  The average time between coarsening-induced rearrangement events at a given scattering site, $\tau_o$, as measured by DWS \cite{DJDpra91}, grows from 40 to 55~seconds across this same age range.  The average duration of events, $\tau_d$, estimated by video of surface bubbles \cite{DJDsci91, AnthonyJCIS99, AnthonyPRL95}, is of order 0.1~seconds.  Optically, bulk samples appear white and hence strongly scatter visible light with negligible absorption.  The transport mean-free path, $l^*$, equal to the typical step size in the random walk taken by photons as they diffuse within the sample, is 3.5 times the average bubble diameter \cite{DJDsci91, MoinAO01}.  The photon paths are not truly random walks, however, because photons are channeled with slight preference along the Plateau borders~\cite{AlexEPL04}.

\subsubsection{Speckle-Visibility Spectroscopy}

\begin{figure}
\includegraphics[width=3.00in]{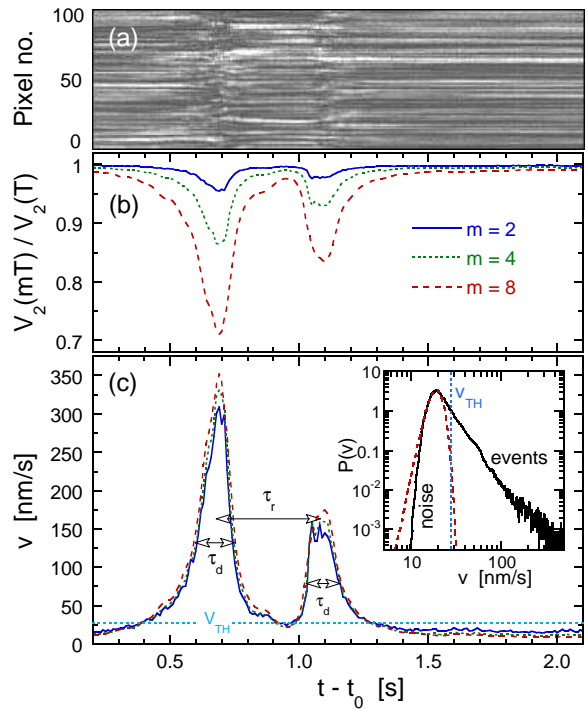}
\caption{(Color online) Example rearrangement events in a coarsening foam, illustrated by time traces of (a) raw pixel grayscale levels, (b) variance ratios computed from video data, and (c) bubble speeds computed from variance ratios.  The foam age is $t_0 = 332$~minutes, and the camera exposure time is $T=10$~ms.  The inset of (c) shows the bubble speed distribution, averaged over age range $326-334$ minutes, with the peak fit to a Gaussian.  The threshold speed for discriminating events from noise is computed from the mode and standard deviation of the Gaussian fit as $v_{\rm TH}=v_{\rm mode}+3\sigma$.  This threshold is indicated by vertical and horizontal dotted lines in the inset and main plot of (c), respectively.  The duration of the events is denoted by $\tau_d$, and the rest time between successive rearrangements is denoted by $\tau_r$, as shown.}
\label{PixV2Events}
\end{figure}

The time-resolved dynamics of bubble rearrangements in the sample are probed by Speckle-Visibility Spectroscopy (SVS).  This method, reviewed in detail in Ref.~\cite{Bandyopadhyay2005}, has been applied to resolve the time-dependent motion of grains subject to periodic vibration~\cite{Dixon2003}, of colloids after cessation of shear~\cite{IanniPRE2006}, and of grains avalanching down a heap~\cite{AbatePRE07}.  Closely related methods include laser-speckle photography \cite{BriersOC81} and Time-Resolved Correlation (TRC) \cite{wongRSI, pineRSI02, LucaJPCM03}.  Here we use the so-called ``point-in / point-out'' illumination / detection geometry, in which the volume of the sample probed by the detected light is not extremely large compared to the rearrangement event volume.  This gives non-Gaussian field statistics for the fluctuations of a single speckle, which invalidates DWS~\cite{GittingsAO2006} but is ideal for SVS.  Specifically, coherent light from a Nd:YAG laser, wavelength $\lambda=532$~nm, is focused across a 3~mm diameter aperture formed by black tape attached and centered on the face of the sample cell.  Photons which re-emerge from this same aperture are collected with a linescan CCD camera (1024 pixels, 8 bits deep), placed such that the speckle size is comparable to the pixel size.  There is no lens, only a line filter to eliminate ambient room light.  The exposure time is set to $T=0.01$~s, and the laser power is adjusted so that the average gray scale level is about 50.  Video data are then collected in 8.5 minute segments, centered at ages \{270, 300, 330, 360, 390, 420\}~minutes, and streamed to disk for post-processing.  This is repeated four times starting with fresh foam, and the results are averaged together for improved statistics; no differences are noticed between the individual runs.

An example space-time plot for a 100 pixel subset of the video data is shown in Fig.~\ref{PixV2Events}(a) for a two second time window at age 332~minutes after injection.  Static speckle is evident at the beginning and end of this window as horizontal randomly-alternating bright and dark streaks; over these times the bubbles are nearly fixed in place.  Two separate bubble rearrangement events happen in between, centered near 0.7 and 1.1 seconds respectively, as evident by a scrambling of the speckle pattern.  For proper choice of exposure time $T$, the speckle is easily visible between events but is smeared and washed out during events.  In SVS motion is thus quantified by the variance of intensity levels, $V_2(T)\propto \langle I^2\rangle_T - \langle I\rangle^2$, where $\langle\cdots\rangle_T$ denotes the average over pixels exposed for duration $T$.  Note that the average intensity is independent of $T$, so no subscript is placed on $\langle I\rangle$.  The proportionality constant of $V_2(T)$ is set by the laser intensity and the ratio of speckle to pixel size.  Both effects may be neatly canceled by considering the variance ratio $V_2(mT)/V_2(T)$, where the numerator is found from a ``synthetic exposure'' equal to the sum of $m$ successive images.  As demonstrated in Fig.~\ref{PixV2Events}(b) for $m=\{2,4,8\}$, this variance ratio equals almost one when the speckle is static and decreases noticeably during rearrangement events -- more so for larger $m$.

The theory of SVS \cite{Dixon2003,Bandyopadhyay2005} may now be invoked to relate the measured variance ratio to the underlying bubble motion.  Specifically, if there is random ballistic speed between adjacent scattering sites with average speed $v$, then the instantaneous power spectrum is Lorentzian with linewidth
\begin{equation}
    \Gamma=(4\pi v/\lambda),
\label{linewidth}
\end{equation}
the corresponding normalized field autocorrelation is $g_1(\tau)=\exp(-\Gamma\tau)$, and the variance ratio is
\begin{equation}
    {V_2(mT)\over V_2(T)} = {e^{-2mx}-1+2mx \over (e^{-2x}-1+2x)m^2 }
\label{SVSratio}
\end{equation}
where $x=\Gamma T$.  From measurements of the variance ratio vs time, we invert this equation to deduce the linewidth and the average bubble speed $v$ vs time.  An example is shown in Fig.~\ref{PixV2Events}(c).  Note that the results for all three values of $m$ give the same bubble speeds, validating our analysis in terms of instantaneous linewidth $\Gamma$.  However we emphasize one caveat: $v$ represents the relative speed of scattering sites spatially averaged over the volume probed by the collected light.  While diffusing photons may wander far into the sample, the primary sampling volume for backscattered photons is approximately the aperture area times $2.5l^*$~\cite{Cox01}. If a bubble rearrangement event is smaller than this scattering volume, then the speed deduced from the linewidth will be an underestimate.  We will present all data in terms of the nominal speeds given by Eq.~(\ref{linewidth}).  As will be discussed, this is accurate for the ``noise'' between events but is small by a factor of nearly $10^4$ for the events.

%=========================================================================================

\section{Telegraph analysis}

\begin{figure}
\includegraphics[width=3.00in]{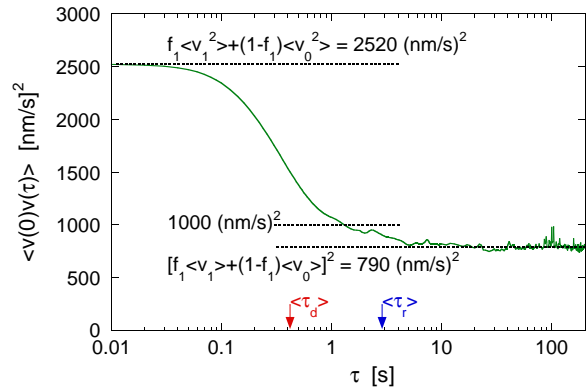}
\caption{Autocorrelation of the velocity time-trace for a coarsening foam over the age range $326-334$ minutes.  The expected short and long time values based on the event and noise statistics of Fig.~\protect{\ref{TimesSpeeds}} are show by dotted lines, as labeled.  The expected intermediate time value, labeled 1000~(nm/s)$^2$, is given by ${f_1}^2\langle v_1\rangle^2 + f_1(1-f_1)\langle v_1\rangle \langle v_0\rangle + (1-f_1)^2\langle {v_0}^2\rangle$.}
\label{Vautocorr}
\end{figure}

In this section we begin analysis of the SVS data on the time-dependence of the average bubble speed, $v(t)$, in terms rearrangement event statistics.  The first step is to identify the events, their durations $\tau_d$, and the rest times $\tau_r$ between events.  These quantities are independent of the connection in Eq.~(\ref{linewidth}) between linewidth and bubble speed.  Since $v(t)$ is small but nonzero between events, as seen for example in Fig.~\ref{PixV2Events}(c), a threshold must be chosen for discriminating events from noise.  For this it is helpful to consider the histogram of bubble speeds, as shown in the inset of Fig.~\ref{PixV2Events}(c) for the age range $326-334$~minutes.  Usually the bubbles are not rearranging, so the speed distribution has a peak at small speeds representing noise, and a long tail at high speeds representing rearrangement events.  A reasonable prescription is to take the threshold as $v_{\rm TH}=v_{\rm mode}+3\sigma$ where $v_{\rm mode}$ is the location of the peak in the speed distribution and $\sigma$ is the width of the peak given by fit to a half-Gaussian for $v \le v_{\rm mode}$.  The result is indicated by a light-blue dashed line labeled $v_{\rm TH}$ in both the main plot and inset of Fig.~\ref{PixV2Events}(c).  Whenever $v(t)$ rises above the threshold, we define the event duration $\tau_d$ as the full-width at halfway between $v_{\rm mode}$ and $v_{\rm max}$.  Compound events are separated out whenever the signal remains above threshold but there is a minimum of depth greater than $3\sigma$ below successive peaks.  Rest times $\tau_r$ are defined as the interval between successive peaks. With these definitions, illustrated graphically in Fig.~\ref{PixV2Events}(c), the integral over an event is well-approximated by $v_{\rm max}\tau_d$.  And the fraction of time spent rearranging is given by the average duration and rest times as $f_1 = \langle \tau_d \rangle / \langle \tau_r \rangle$.

The integrity of these procedures may be tested using the autocorrelation of $v(t)$, shown for example in Fig.~\ref{Vautocorr} for the same $326-334$~minute age range as Fig.~\ref{PixV2Events}.  Features are to be compared with expectations based on considering $v(t)$ as a telegraph signal that switches between rest=0 and event=1 levels with statistics given by the values of \{$\langle \tau_d \rangle$, $\langle \tau_r \rangle$, $f_1 = \langle \tau_d \rangle / \langle \tau_r \rangle$\} plus the mean and mean-squared speeds \{$\langle v_0 \rangle$, $\langle {v_0}^2 \rangle$, $\langle v_1 \rangle$, $\langle {v_1}^2 \rangle$\} computed respectively from $P(v)$ with $v$ below/above threshold.  First note that the speed autocorrelation exhibits a two-step decay, with the first located near the average event duration $\langle \tau_d \rangle$ and with the second located near the average rest time $\langle \tau_r \rangle$.  At short times, the autocorrelation is very close to $f_1\langle {v_1}^2 \rangle + (1-f_1) \langle {v_0}^2 \rangle$.  At long times, the autocorrelation is very close to $[f_1\langle {v_1} \rangle + (1-f_1) \langle {v_0} \rangle]^2$.  At intermediate times, between the two decays, there is an inflection near ${f_1}^2\langle v_1\rangle^2 + f_1(1-f_1)\langle v_1\rangle \langle v_0\rangle + (1-f_1)^2\langle {v_0}^2\rangle$.  For a true telegraph signal this plateau would have been flatter.  But the clear signature of all these features in the speed autocorrelation shows that we have achieved satisfactory identification of events.

Scatter plots (not shown) of quantities \{$\tau_{d,i+1}$, $\tau_{r,i+1}$, $v_{{\rm max},i+1}$\} pertaining to event $i+1$ vs quantities \{$\tau_{d,i}$, $\tau_{r,i}$, $v_{{\rm max},i}$\} pertaining to the previous event $i$ reveal no evidence of correlations.  Thus the switching between rest and event states appears to be Markovian, proceeding randomly at a history-independent rate.

%=========================================================================================

\section{Evolution of averages}

\begin{figure}
\includegraphics[width=3.00in]{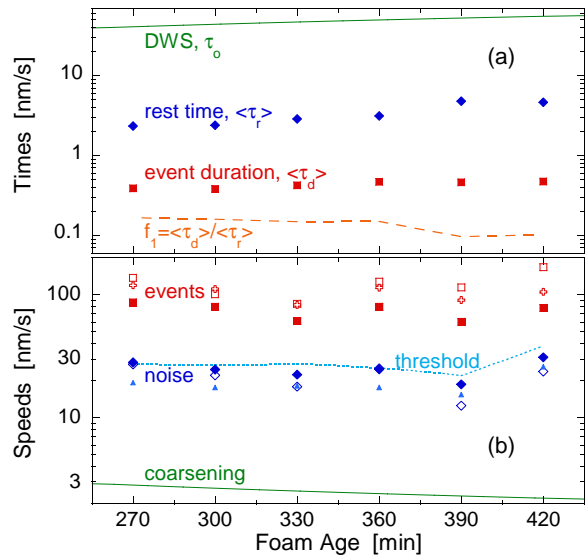}
\caption{(Color online)  Characteristic time and speed scales vs foam age.  In (a) the solid diamonds represent the rest time between successive rearrangement events in the whole scattering volume and the solid squares represent the event durations, both averaged over 30 minute runs; the fraction of time the bubbles in the whole scattering volume are in motion is shown by the dashed curve $f_1$.  The time $\tau_o$ between successive events at one scattering cite was previously measured by diffusing-wave spectroscopy (DWS) and is shown by the solid curve.   In (b) statistics for event speeds are shown by solid squares for the average, open squares for the standard deviation, and open pluses for the average of the maxima.  Statistics for noise speeds are shown by solid diamonds for the average, open diamonds for the standard deviation, solid triangles for the mode, and dotted curve for the threshold; the latter two are from Gaussian fits to the full speed distribution as shown in the inset of Fig.~\protect{\ref{PixV2Events}}(c).  The rate of change of the average bubble diameter, due to coarsening, was previously measured and is shown by the solid curve labeled ``coarsening''.}
\label{TimesSpeeds}
\end{figure}

\begin{figure}
\includegraphics[width=3.00in]{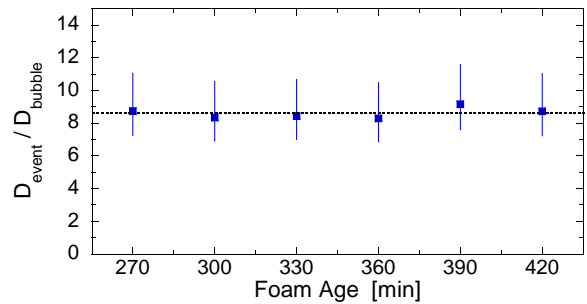}
\caption{(Color online) Ratio of average rearrangement event diameter to bubble diameter vs foam age, based on SVS data for the average time $\langle \tau_r \rangle$ between events in the scattering volume and DWS data for the time $\tau_0$ between events at a given scattering site.  The average event size is 8.6 bubble diameters across for all foam ages (horizontal line).  Error bars represent the range of results taking the scattering volume as aperture area times $(2.5\pm1.5)l^*$.}
\label{Devent}
\end{figure}

The average statistical quantities extracted from telegraph analysis of the SVS speed signals are displayed vs foam age in Fig.~\ref{TimesSpeeds}.  The average rest times and event durations are shown in the top plot, along with their ratio $f_1$ and the DWS results for the time $\tau_o$ between successive events at the same scattering site.  There is good separation of time scales, $\tau_o \gg \langle \tau_r \rangle \gg \langle \tau_d \rangle$.  The durations are constant for all ages studies; the average value is $\langle\langle \tau_d \rangle\rangle=0.44$~s in accord with video observation of surface bubbles.  The rest times grow longer with age, like $\tau_o$, but not dramatically; the average value is $\langle\langle \tau_r \rangle\rangle=2.9$~s.  Note that these results for event and rest times are independent of any underestimation in bubble speeds due to inapplicability of Eq.~(\ref{linewidth}) from interpreting localized motion as an average over the scattering volume.

The key speed statistics are shown in the bottom plot of Fig.~\ref{TimesSpeeds}.  For the events, this includes the average and standard deviation during events, and the average peak speed; these quantities are all approximately independent of foam age.  As expected since the speed during an event rises and falls, the standard deviation is greater than the average.  Thus, in accord with a telegraph analysis, the best quantity to consider is the peak speed; the average over all ages is $\langle \langle v_{\rm max} \rangle \rangle = 120$~ns/s.  For the ``noise'' speed statistics, the plotted values include the average, the standard deviation, the mode, as well as the threshold used to discriminate events.  These values are all comparable, are roughly constant~\cite{Veronique}, and are much smaller than the event speeds.  We also observe in Fig.~\ref{TimesSpeeds} that the noise speed is ten times larger than the rate of change of the average bubble size, ${\rm d}\langle D_{\rm bubble}\rangle/{\rm d}t$, caused by coarsening.  Therefore we infer that the noise in the SVS signal is due to the nonaffine relative motion of bubbles and plateau borders, continuously adjusting their positions in response to gradual size changes in order to optimize packing for locally-minimal surface area.  While such motion maintains each bubble at the bottom of an energy well, the wells can eventually become unstable and produce a sudden rearrangement event.

The average time $\langle \tau_r  \rangle$ between events in the scattering volume can be understood in terms of the number of bubbles in an event.  Since $\tau_r$ involves the whole scattering volume and the DWS time $\tau_o$ involves only a single event volume, the value of $\tau_r$ is smaller than $\tau_o$ according to the ratio of event to scattering volumes: $\tau_r = (V_{\rm event} / V_{\rm scattering}) \tau_o$.  Taking the scattering volume as aperture area times $2.5l^*$ \cite{Cox01}, and taking the transport mean free path $l^*$ as 3.5 times the average bubble diameter \cite{DJDsci91, MoinAO01}, we thus deduce the ratio of event to bubble diameter and plot the value vs age in Fig.~\ref{Devent}.  The results are nearly constant at $D_{\rm event}/D_{\rm bubble}=8.6$ independent of foam age.  Though clearly finite, unlike avalanches which propagate out to the system size, this is still large compared to the bubble size.  Video of surface rearrangements suggest that only a small core of neighboring bubble actually undergo topology change.  Hence there is a halo of non-affine motion surrounding this ``point-like'' perturbation so that the bubbles maintain a locally optimal packing.  The extent of this reaction decays with distance, and falls below one wavelength at about 8.6 bubble diameters.  Fluctuations in the TRC signal are also related to event size, though the precise connection and the resulting event size were not reported \cite{VeroniquePRL04}.  See Ref.~\cite{AdamPRE07b} for an explicit discussion of how to deduce the size of dynamic heterogeneities from the variance in the decay of an autocorrelation.

%=========================================================================================

\section{Time distributions}

\begin{figure}
\includegraphics[width=3.00in]{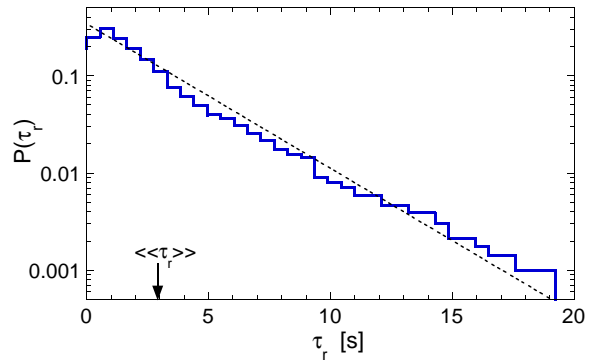}
\caption{Probability distribution for the rest time between successive bubble rearrangements in the scattering volume, averaged over all ages.  For all events combined, the average is indicated by $\langle \langle \tau_r \rangle \rangle=2.9$~s.  The dashed line represents an exponential distribution with this average.}
\label{PTrest}
\end{figure}

\begin{figure}
\includegraphics[width=3.00in]{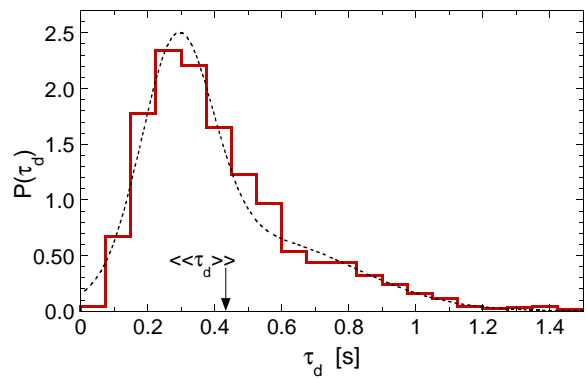}
\caption{Probability distribution for the duration of individual events, averaged over all ages.  The dashed curve represents the best fit to the sum of two Gaussians, with respective peaks at 0.29 and 0.54~s, with widths of 0.11 and 0.28~s, and with weights of 0.55 and 0.45. For all events combined, the average is indicated by $\langle \langle \tau_d\rangle\rangle=0.44$~s.}
\label{PTevent}
\end{figure}

One advantage of SVS and the telegraph analysis is that the full distribution of all times and speeds are easily tabulated.  For example, in Fig.~\ref{PTrest} we display the distribution $P(\tau_r)$ of rest times between events on semilogarithmic axes.  Since the average values considered above are nearly independent of time, $\langle \langle \tau_r \rangle \rangle=2.9$~s, we combine together the results for all ages.  The results exhibit a form which is nearly exponential, as demonstrated by comparison with the straight line representing $\exp[-\tau_r/(2.9~{\rm s})]/(2.9~{\rm s})$.  This provides further support for the Markovian nature of the rearrangements noted above.  Rearrangements occur seemingly at random with a given rate, just like the radioactive decay of unstable nuclei.  The new packing states just after rearrangement must have a range of well depths.

Next the full distribution $P(\tau_d)$ of the duration of events is displayed in Fig.~\ref{PTevent}.  Again, since the averages are nearly constant, $\langle \langle \tau_d \rangle \rangle =0.44$~s, we combine the results for all ages.  The form of the distribution rises abruptly from zero, peaks narrowly between 0.2~s and 0.4~s, then decays more gradually out to about 1.2~s.  Because the rest times are exponentially distributed with an average only about ten times longer than the event duration, and because the scattering volume is larger than the event volume by the factor $\tau_o/\langle \tau_r \rangle=17$, it cannot be uncommon for two unrelated events to occur simultaneously.  Therefore we fit $P(\tau_d)$ to a sum of two Gaussians.  The result shown by a dashed curve describes the behavior fairly well.  The first peak is centered at 0.29~s, has width 0.11~s, and weight 0.55; this would represent singly-occurring events.  The second peak is centered at 0.54~s, has width 0.28~s, and weight 0.45; this would represent multiple events.

The average duration of events is a crucial microscopic quantity for understanding foam rheology, because during flow the nature of bubble motion changes dramatically according to whether the shear rate is high or low compared to the reciprocal of event duration \cite{ParkPRL94, AnthonyJCIS99, AnthonyPRL03}.  We note that the value found here, $\langle \langle \tau_d \rangle \rangle =0.44$~s, is comparable to the relaxation time of a single Plateau border observed in a quasi-2D foam made with sodium-dodecylsulfate \cite{DurandStone06}.  This time scale was argued to set by the sum of surface shear and dilatational viscosities divided by surface tension, independent of bubble size and bulk shear viscosity.  In support, note in Fig.~\ref{TimesSpeeds} that the event durations are independent of age even though the bubble radii and the time between events both grow noticeably.

%=========================================================================================

\begin{figure}
\includegraphics[width=3.00in]{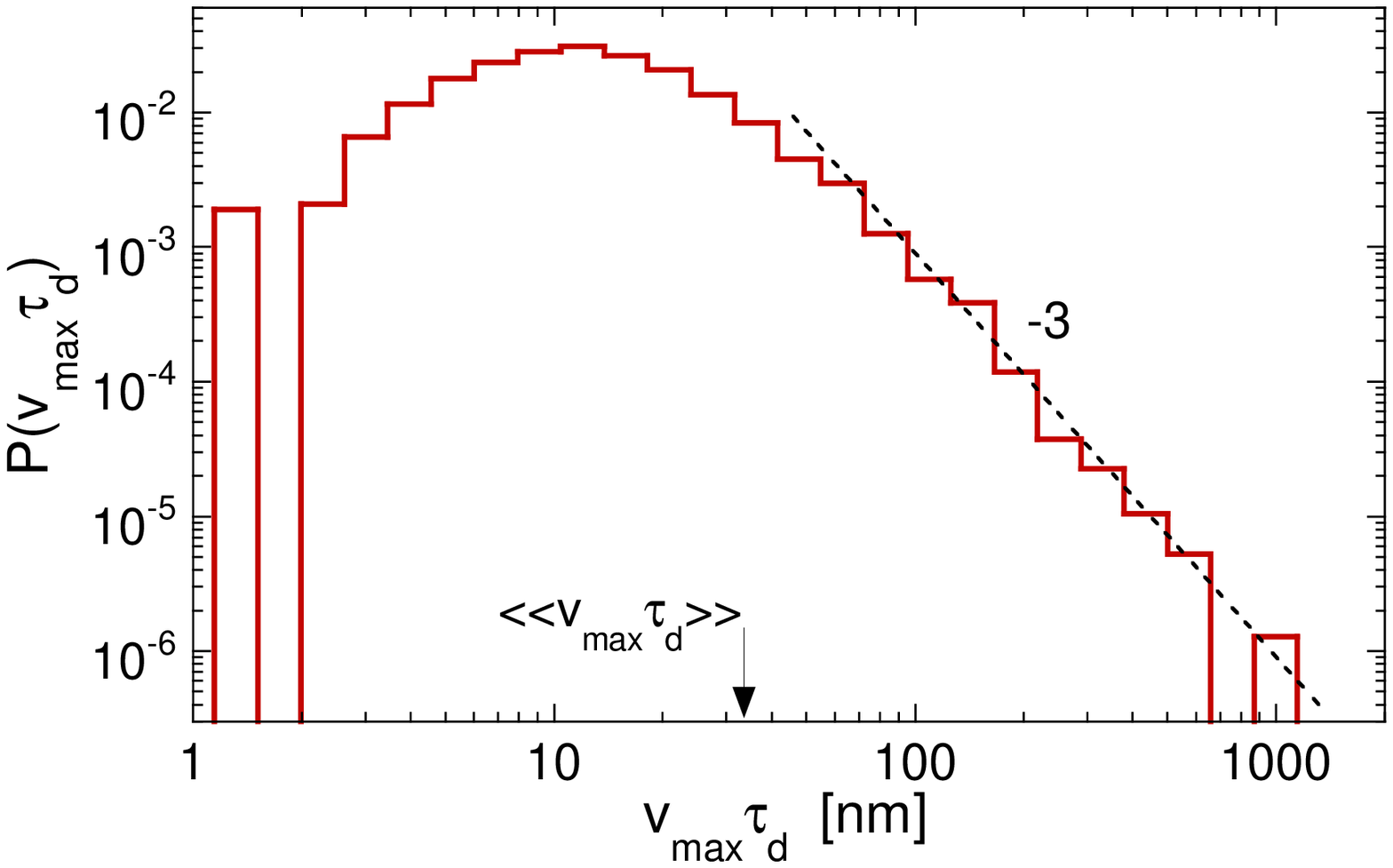}
\caption{Probability distribution for the event displacements, $v_{\rm max}\tau_d$, averaged over all foam ages.  The dashed line represents fit to a power-law tail with exponent as labeled.  For all events combined the average is indicated by $\langle \langle v_{\rm max}\tau_d \rangle\rangle$.}
\label{Pdisp}
\end{figure}

\begin{figure}
\includegraphics[width=3.00in]{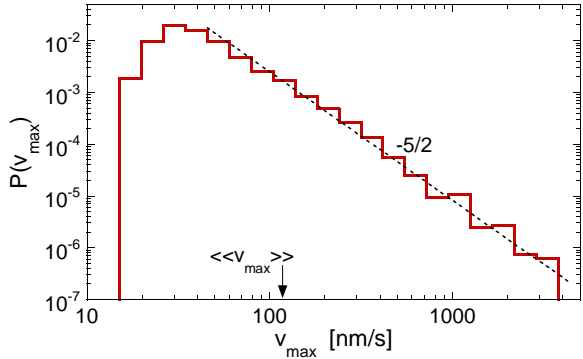}
\caption{Probability distribution for peak event speeds, averaged over all ages.  The dashed line represents fit to a power-law tail with exponent $-5/2$ as labeled.  For all events combined the average is indicated by $\langle \langle v_{\rm max} \rangle\rangle$.}
\label{PVmax}
\end{figure}

\begin{figure}
\includegraphics[width=3.00in]{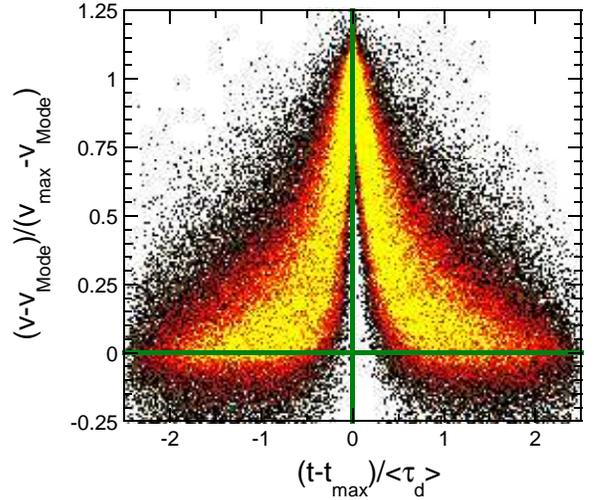}
\caption{Probability density map for the speed vs time during an event; the color code for most to least probable is yellow, orange, red, maroon, brown, black.  The speed axis is offset by the mode of the speed distribution and normalized by peak speed for each event; the time axis is offset by the time at which the speed is maximum and normalized by the event duration of each event.  Only events in the power-law tail of the peak speed distribution of Fig.~\protect{\ref{PVmax}} are included.}
\label{BinEvents}
\end{figure}

\section{Motion distributions}

In this section we consider the nature of the bubble motion during events.  By the definition of event durations as the full-width half-max of peaks in the $v(t)$ signal, the displacement due to an event is well-approximated by $v_{\rm max}\tau_d$.  The distribution of all such displacements, for all runs and all ages, is plotted on logarithmic axes in Fig.~\ref{Pdisp}.  It exhibits a peak around 12~nm, an average value of $\langle \langle v_{\rm max}\tau_d \rangle\rangle=32$~nm, and a long power-law tail with exponent $-3$.  It is difficult to precisely interpret these results because, as noted above, the connection between the SVS linewidth and bubble speed assumes that each scattering site has the same dynamics whereas the actual rearrangements are spatially localized.  Furthermore, the bubble motion itself is heterogenous, consisting of a core undergoing neighbor-switching and a halo undergoing nonaffine adjustment at fixed topology that decays in amplitude away from the core.  Both effects may contribute to causing the apparent power-law tail.  And certainly both effects contribute to making $v(t)$ an underestimate of the bubble speed in the core.  If we suppose that $v(t)$ is dominated by the core motion, then we can correct for the underestimate by multiplying by (1) the ratio $\tau_o/\langle\tau_d\rangle\approx 50/0.44$ of scattering to event volume and (2) the ratio $(8.6/2)^3$ of halo to core volume; this product equals 9000, the number of bubbles in scattering volume.  The peak in $P(v_{\rm max}\tau_d)$ would then shift by a factor of 9000 from 12~nm to 110~$\mu$m.  This is comparable to the average bubble diameter, which is the expected scale of displacement in the core.

Further support for the above correction to the inferred speeds comes from consideration of the distribution of peak bubble speeds, shown on logarithmic axes Fig.~\ref{PVmax}.  Again this distribution is a compilation of results for all runs and all ages.  It exhibits a peak around 30~nm/s, an average value of $\langle \langle v_{\rm max}\rangle\rangle=120$~nm/s, and a long power-law tail with exponent $-5/2$.  Applying the same factor of 9000 shifts the peak in the maximum speed distribution from 30~nm/s to 270~$\mu$m/s.  This value is in good agreement with an estimate of speed as equal to bubble diameter divided by event duration, $v=\langle D_{\rm bubble}\rangle / \langle \tau_d \rangle = 250$~$\mu$m/s.  Thus the analysis is self-consistent.

Lastly we inspect the typical profile of speed vs time during an event in Fig.~\ref{BinEvents}.  Since there are broad distributions of event durations and peak speeds, we normalize each event separately according to these characteristics and compile the results.  To reduce the scatter, we also subtract the mode of the speed distribution, and include only events in the power-law tail of the peak speed distribution, which, according to Fig.~\ref{PVmax}, begins at nominal speed of 50~nm/s.  The color coding of bins in Fig.~\ref{BinEvents} is such that the most probable event shape is traced out along the crest of the yellow region.  This reveals that events begin and end rather gradually, but speed up and slow down dramatically on either side of the peak.  The event shapes are nearly symmetrical around the peak, with the rise being only somewhat faster than the decay.  The first event in Fig.~\ref{PixV2Events} is typical in its sharpness, while the second is more typical in its slight asymmetry.  It would be interesting to know where the actual topology change occurs in relation to the peak speed.  Perhaps the bubble accelerate toward a neighbor change which happens at peak speed, or perhaps the bubbles remain at rest until motion is initiated by a neighbor change.

%=========================================================================================
\section{Conclusion}

For a very large scattering volume of foam, for instance with a ``plane-in / plane-out'' illumination/detection geometry, multiple rearrangement events are present and each causes only slight speckle change since most light paths are unaffected.  Thus the speckle fluctuations are nearly continuous and exhibit Gaussian field statistics at a single far-field detection site~\cite{GittingsAO2006}.  This is the usual requirement for intensity-correlation spectroscopy versions of dynamic light scattering like Diffusing-Wave Spectroscopy (DWS).  Under these conditions analysis of DWS data yields the rate of rearrangement events for coarsening and slowly sheared foams, or the bubble speed for rapidly sheared foams.  If the scattering volume is small enough so that often no rearrangements are present, and so that many light paths are affected by single events, then non-Gaussian field statistics in time-averaging at a single detection site will prevent application of DWS~\cite{GittingsAO2006}.  Under these conditions, there is extra information that can be accessed either by higher-order time-averaged intensity correlations~\cite{PierreJOSA99, GittingsAO2006} or by time-resolved multispeckle methods that ensemble-average over the speckle pattern.

In this paper we probed coarsening-induced bubble dynamics in a small scattering volume with Speckle-Visibility Spectroscopy (SVS), which is a particularly simple and fast multispeckle dynamic light scattering method.  The visibility of the speckle pattern in a single exposure of a CCD camera was repeatedly measured, then converted to a time trace of the linewidth of the scattered light.  This signal exhibited a sequence of spikes representing individual discrete rearrangement events.  Thus it was straightforward to identify the duration of events and of the rest times between events, which we analyzed in terms of both moments and full distributions.  From lack of history dependence and from the exponential distribution of rest times, we concluded that rearrangement events are Markovian.  From comparison of rest times with previous DWS data for the time between events a a single scattering site, we concluded that the event size is almost nine bubble in diameter.  Events consist of a small core of bubbles that switch neighbors, surrounded by a halo that shift at fixed topology.

It was less straightforward to analyze the magnitude of the linewidth in terms of bubble speeds.  Between events, the only motion is due to slight bubble size changes and gradual adjustment of bubble positions and Plateau borders to the changing packing conditions.  This is stochastic but homogeneous both temporally and spatially throughout the scattering volume, and hence was readily analyzed in terms of bubble speeds.  However, rearrangements events are spatiotemporally heterogeneous so that the usual prescription for conversion of linewidth to speed gives a severe underestimate.  We argued that the correction factor is set both by the ratio of scattering to event volume and the ratio of event to core volume.  We showed that this gives both the correct total displacement within the core, and the correct speed within the core.  In future studies, it would thus be a fine procedure to apply a scale factor to the measured linewidth so that the resulting typical displacement equals the bubble diameter.

While the contributions here advance the understanding of both coarsening and rearrangement dynamics, and how they may be probed by time-resolved dynamic multiple-light scattering, there are many open lines of research.  For example, it would be interesting to systematically vary the scattering volume, both as a further check on the procedures presented here and also perhaps as a way of spatially-resolving the motion in terms of a diffuse-light tomography.  Relatedly it would be interesting to consider the effect of the location of an event on the SVS signal.  Also for example, it would be interesting to vary the aqueous solution in attempt to control and understand rearrangement rates and durations.  And finally, it would be interesting to understand the non-affine shifts in response to coarsening or slow shear that ultimately lead to rearrangements.

\begin{acknowledgments}
We thank L.~Cipelletti, R.~H\"ohler, and V.~Trappe for helpful conversations. This work was supported by NASA Microgravity Fluid Physics Grant NNX07AP20G.
\end{acknowledgments}

% figure list
% 1 PixV2Events
% 2 TimesSpeeds
% 3 Vautocorr
% 4 Devent
% 5 PTrest
% 6 PTevent
% 7 PVmax
% 8 Pdisp
% 9 BinEvents

%
% To reduce eps size:
% [1] open with Ghostscript and convert/save as PDF 300
% [2] open with Adobe and save as PS
% [3] open with Ghostscript and Convert PS to EPS

\bibliography{SvsCoarseningFoams_refs}

\end{document}